\documentclass[reprint,aps,superscriptaddress,amsmath,amssymb]{revtex4-2}

\usepackage{graphicx}
\usepackage{dcolumn}
\usepackage{bm}
\usepackage{color}
\usepackage[dvipsnames]{xcolor}
    \definecolor{LST}{rgb}{0.1, 0.7, 0.1}
    \definecolor{marked_text}{rgb}{0, 0, 0}
\usepackage{placeins}

\usepackage[normalem]{ulem}

\usepackage[linkcolor=red,urlcolor=blue,citecolor=blue,colorlinks=true]{hyperref}

\begin{document}

\title{Generation of intense, polarization-controlled magnetic fields \\with non-paraxial structured laser beams}

\author{Sergio Martín-Domene}
\email{sergiomardom@usal.es}
\affiliation{Grupo de Investigación en Aplicaciones del Láser y Fotónica, Departamento de Física Aplicada, Universidad de Salamanca, E-37008, Salamanca, Spain}
\affiliation{Unidad de Excelencia en Luz y Materia Estructuradas (LUMES), Universidad de Salamanca, Salamanca, Spain \looseness=-1 } 

\author{Luis Sánchez-Tejerina}
\affiliation{Departamento de Electricidad y Electrónica, Universidad de Valladolid, 47011, Valladolid, Spain \looseness=-1 }

\author{Rodrigo Martín-Hernández}
\affiliation{Grupo de Investigación en Aplicaciones del Láser y Fotónica, Departamento de Física Aplicada, Universidad de Salamanca, E-37008, Salamanca, Spain}
\affiliation{Unidad de Excelencia en Luz y Materia Estructuradas (LUMES), Universidad de Salamanca, Salamanca, Spain \looseness=-1 }

\author{Carlos Hernández-García}
\affiliation{Grupo de Investigación en Aplicaciones del Láser y Fotónica, Departamento de Física Aplicada, Universidad de Salamanca, E-37008, Salamanca, Spain}
\affiliation{Unidad de Excelencia en Luz y Materia Estructuradas (LUMES), Universidad de Salamanca, Salamanca, Spain \looseness=-1 }

\begin{abstract}
The ability to spatially separate the electric and magnetic fields of a light beam enables the inspection of laser-matter interactions driven solely by optical magnetic fields. However, magnetic field excitations are commonly orders of magnitude weaker than those driven by the electric field. Several studies have already demonstrated the isolation of an intense, linearly polarized magnetic field using structured light. In this work, we report the generation of isolated high intensity magnetic fields with controlled polarization state in the non-paraxial regime using structured laser beams. Our theoretical findings highlight a significant enhancement in the amplitude of the longitudinal magnetic field carried by an azimuthally polarized laser under tight-focusing conditions. Furthermore, by implementing a multiple-beam configuration, we achieve precise control over the polarization state and amplitude of the spatially isolated magnetic field. We report the generation of polarization-controlled magnetic fields reaching up to tens of Tesla, even from moderately intense laser beams of $\sim 10^{12} \, \mathrm{W}/\mathrm{cm}^2$. Our study paves the way for ultraintense interactions with circularly polarized magnetic fields from a feasible experimental setup point of view, particularly interesting to probe ferromagnetic materials and chiral media. 
\end{abstract}

\maketitle


According to Maxwell's equations, 
electric (E-field) and magnetic (B-field) fields are inherently coupled. 
The most straightforward solution to the wave equation is the well-known plane wave, where the fields exhibit spatially homogeneous amplitude, polarization state and phase along a plane perpendicular to the propagation direction \cite{Burko_2008}. More intricate electromagnetic field distributions, featuring complex spatial variations in amplitude, phase, and polarization state, can be derived from the Helmholtz equation. These solutions give rise to what is commonly known as structured light \cite{A.Forbes_structured_laser_review, 2022_He_structured}. In recent years, structured light beams have unlocked diverse possibilities across various fields \cite{Roadmap_structured_light, 2023_bliokh_Roadmap, 2023_shen_Roadmap}, including optical manipulation, optical communications, quantum technologies, ultrafast science and condensed matter systems. The most common structured beams, exhibiting spatially homogeneous polarization states, can be solved analytically within the paraxial approximation. Examples include Hermite-Gauss, Laguerre-Gauss, and Ince-Gauss solutions, which are found in cartesian, cylindrical, and elliptical coordinates, respectively \cite{Kogelnik_scalar_modes,IG_Bandres}.

By solving the Helmholtz equation in its vectorial form within the paraxial approximation, cylindrical vector beams (CVBs) can be obtained \cite{Hall_CVB_solutions}, characterized by their spatially non-homogeneous polarization state. Notable examples include radially and azimuthally polarized beams, i.e. linearly polarized beams with a varying tilt-angle along the transverse plane, depicting a radial or azimuthal pattern, and exhibiting a singularity at the beam axis. These beams have been used in applications in diverse disciplines such as material procesing \cite{drevinskas2016_procesing,Pallares2023_procesing}, particle trapping \cite{Moradi2019_part_trapping} and acceleration \cite{Salamin2010_part_accel,Xu2016_part_accel}, microscopy \cite{Liu2022_microscopy} or quantum optics \cite{Watzel2019_quantum_opt,Parigi2015_quantum_opt}. A particularly interesting property of azimuthally polarized beams is that they present a longitudinally polarized B-field on-axis, precisely at the E-field singularity. This feature has boosted the interest in applications driven solely by the interaction with the locally isolated B-field. Examples include studies in magnetic spectroscopy \cite{Veysi2016, 2021_wozniak}, force microscopy \cite{Zeng2018}, optical spectroscopy \cite{2015_kasperczyk}, or ultrafast magnetization dynamics \cite{Luis_Sanchez-Tejerina2023}. The increased interest in these unique magnetic probes has prompted the exploration of their diverse applications. However, the efficiency of matter interactions driven by B-fields is orders of magnitude weaker than those driven by E-fields. Thus, enhancing the intensity of such isolated B-fields is highly desirable \cite{2018_sanz-paz}. 
Various approaches that make use of azimuthally polarized laser beams with peak intensities of $\sim 10^{11}-10^{14} \, \mathrm{W/cm}^2$ have explored the obtention of B-fields above the Tesla level. In particular, by inducing electrical currents in metallic \cite{M.Blanco_B_pulse} or gaseous \cite{Sederberg2020_B_pulse} media, ultrafast Tesla B-fields have been proposed. More recently, tailored nanoantennas have been discussed to further enhance such longitudinal B-fields \cite{Rodrigo_Lorenz}. All these approaches make use of the linearly polarized longitudinal B-field carried by an azimuthal beam. However, this feature excludes applications in which the control over the B-field polarization state is crucial, such as to drive nonlinear magnetization dynamics in ferromagnets \cite{Luis_Sanchez-Tejerina2023}, or to explore chiral media \cite{magnetic_chirality_Cheong2022,Tang2010}.

In this paper we theoretically explore the generation of isolated intense B-fields with controlled polarization states by using tightly-focused azimuthally polarized beams. We first study the B-field enhancement in the non-paraxial regime using the Richards-Wolf vectorial diffraction method \cite{RW1, RW2}, revealing significantly higher isolated B-fields than in the paraxial approximation. Subsequently, we introduce a technique to generate an intense and isolated B-field with a tunable polarization state, ranging from linear to circular. The proposed method involves the coherent addition of tightly focused azimuthally polarized beams in a crossed geometry. Our research work does not only provide insights into achieving higher B-fields beyond the paraxial approximation, but also paves the way for intense and selective B-field interactions where the polarization state plays a crucial role.


Azimuthally polarized laser beams present an E-field along the azimuthal direction, $\mathbf{E} = E_{\phi} \, \hat{\mathbf{u}}_{\phi}$, and a B-field along the radial and longitudinal directions, $\mathbf{B} = B_r \, \hat{\mathbf{u}}_r + B_z \, \hat{\mathbf{u}}_z$. These CVBs are often modeled in the paraxial approximation with an amplitude profile corresponding to that of the first-order Laguerre-Gaussian mode without the azimuthally dependent vortex phase term, \textcolor{marked_text}{solution of the Helmholtz equation.} At the beam waist ($z=0$), the electromagnetic field in cylindrical coordinates then takes the form \cite{Veysi2015_Bz}
\begin{equation}\label{parax_E}
    \mathbf{E}(r) = E_0 \, \frac{r}{w_0} \, e^{-r^2 / w_0^2} \, \hat{\mathbf{u}}_{\phi} ,
\end{equation}
\begin{eqnarray}\label{parax_B}
     \mathbf{B}(r) =&& - \frac{E_0}{c}\, \frac{r}{w_0} \, e^{-r^2 / w_0^2} \, \hat{\mathbf{u}}_r \nonumber\\
     && - i \, \frac{E_0}{c}\, \frac{\lambda}{\pi w_0} \left( 1 - \frac{r^2}{w_0^2} \right) \, e^{-r^2 / w_0^2} \, \hat{\mathbf{u}}_z,
\end{eqnarray}
being $E_0$ the E-field amplitude, $\lambda$ the wavelength, 
and $w_0$ the beam waist. In this work, the attention will be focused on the longitudinal component of the B-field, $B_z$. \textcolor{marked_text}{The use of higher-order CVBs does not provide better enhancement of $B_z$ (see Section 1 of the Supplementary Material).}
Note that $B_r \gg B_z$ in the paraxial approximation ($w_0 \gg \lambda$). Within this regime, upon focusing with an ideal optic system of focal length $f$, the beam waist is reduced to $w_0' = w_0 / \sqrt{1 + (z_R/f)^2}$, being $z_R= \pi w_0^2 /\lambda$ the Rayleigh distance. For tight-focusing conditions---small focal length, $f$, or high numerical-aperture (NA) systems---, the previous formula predicts a vanishing beam waist, so the paraxial approximation is no longer valid. In order to discern between this two regimes, we introduce the F-number or focal ratio $F_{\#} = f/2w_0$, a dimensionless magnitude expressing the relationship between the focal length and the entrance pupil diameter of an optical system. \textcolor{marked_text}{The latter is usually taken as the beam diameter when the pupil plane is placed at $z=0$ as in our configuration.} This way, $F_{\#} \lesssim 1$ indicates the situation of tight-focusing conditions, while $F_{\#} \gg 1$ in the paraxial regime.

\begin{figure}
\centering
\includegraphics[width=8.5cm]{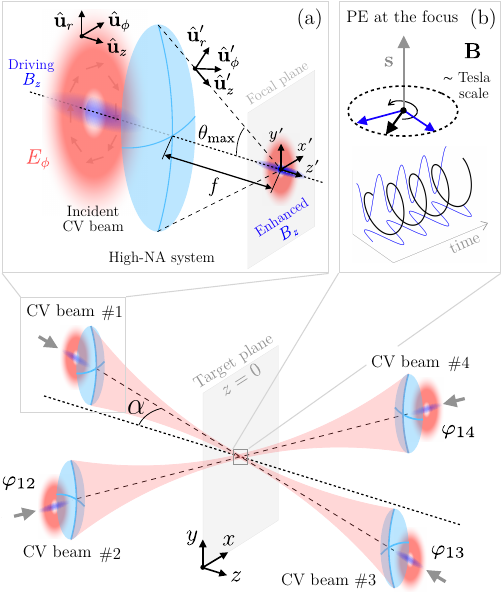}
\caption{Scheme for the generation of polarization controlled B-fields from the coherent addition of azimuthally polarized laser beams. Main panel: four-beam configuration propagating in two pairs of orthogonal directions with the focus placed at the common reference system origin. The B-field is characterized at the target $xy$-plane. The reference beam for the relative phases is the one labeled as \#1 \textcolor{marked_text}{and the observation angle $\alpha$ defines its propagation angle with respect to the $z$-direction, perpendicular to the target plane.} (a) Tight-focusing scheme for a single incident azimuthal CVB in cylindrical coordinates ($B_r$ component not shown). (b) Conceptual picture of the circularly polarized B-field achieved at the focus and its associated polarization ellipse.}
\label{focusing_schemes}
\end{figure}

The proper description of tightly-focused CVBs in the non-paraxial regime is given by the Richards-Wolf vectorial diffraction theory. \textcolor{marked_text}{For the previous azimuthally polarized beam in Eq.~\eqref{parax_E}, the focused electromagnetic field with a high-NA optical system placed at its waist---see Fig.~\ref{focusing_schemes}(a)---is described in terms of the pupil apodization function $P(\theta)$} by \cite{Zhan_CVB_review}
\begin{equation}\label{non-parax_Ephi}
    E_{\phi}(r,z) = A \int_0^{\theta_{\mathrm{max}}} P(\theta) \sin \theta \, J_1(kr \sin \theta)\, e^{ikz\cos \theta} \mathrm{d} \theta ,
\end{equation}
\begin{eqnarray}\label{non-parax_Br}
    B_r(r,z) =&& \, -\frac{A}{c} \int_0^{\theta_{\mathrm{max}}} P(\theta) \sin \theta \cos \theta \, J_1(kr \sin \theta) 
    \nonumber\\
     && \times \, e^{ikz\cos \theta} \mathrm{d} \theta ,
\end{eqnarray}
\begin{eqnarray}\label{non-parax_Bz}
    B_z(r,z) =&& \, -\frac{iA}{c} \int_0^{\theta_{\mathrm{max}}} P(\theta) \sin^2 \theta \, J_0(kr \sin \theta) 
    \nonumber\\
     && \times \, e^{ikz\cos \theta} \mathrm{d} \theta ;
\end{eqnarray}
where we define the amplitude factor $A \equiv 2 \pi f E_0 / \lambda$.
\textcolor{marked_text}{The aplanatic optical system obeys the sine condition, so $P(\theta)$ is given by the product of the obliquity factor $\sqrt{\cos \theta}$ and the paraxial spatial profile of the input E-field in Eq. \eqref{parax_E}, $E_{\phi}(\theta)/E_0 = (f \sin \theta / w_0) \, e^{- ( f \sin \theta / w_0)^2}$ (besides a normalization constant), where we have used that $r~=~f \sin \theta$ at the lens plane.}
$J_{\nu}(\cdot)$ is the first kind Bessel function of order $\nu$; and $k = 2 \pi /\lambda$ is the wavenumber. 
In air, the maximum aperture angle $\theta_{\mathrm{max}}$ with respect to the optical axis is given by $\mathrm{NA} = \sin \theta_{\mathrm{max}}$. Approximately, $\theta_{\mathrm{max}} \approx \arctan (1/ 2 F_{\#})$. 

In order to precisely adjust the polarization state of the B-field, we propose a configuration in which several azimuthally polarized laser beams are coherently arranged, exploiting the fact that their sum is a valid solution of Maxwell equations. 
By tuning their amplitude or phase relationship, we can define an overlapping region where we can control the spatial distribution of the polarization state of the isolated B-field at a given target plane, 
as sketched in Fig.~\ref{focusing_schemes}. 
We shall consider two and four azimuthally polarized beams arranged orthogonally, i.e. whose propagation directions differ by $90^{\circ}$. We fix the target position along the $xy$-plane ($z=0$). \textcolor{marked_text}{The observation angle $\alpha$ defines the propagation direction of the reference beam \#$1$ with respect to the $z$-direction.} The relative phases between each beam and the reference beam \#$1$ are given by $\varphi_{1j}$ for $j=2,3,4$.

\textcolor{marked_text}{The characterization of the local polarization for tridimensional electromagentic fields as our tightly-focused beams have already been addressed both theoretically and experimentally \cite{exp_3d_polariz}.} We opted to characterize the polarization state of the resulting B-field using a non-paraxial formalism \cite{M.Alonso_nonparax_polariz} in which the polarization ellipse (PE) is defined at each point in space, changing its shape and orientation locally along the target plane---see Fig.~\ref{focusing_schemes}(b). The local polarization state description through the PE ellipticity, $\varepsilon$, allows to discriminate between linear ($\varepsilon = 1$), circular ($\varepsilon = 0$), or elliptical polarization states  ($0 < \varepsilon < 1$). \textcolor{marked_text}{The so-called spin density represents the normal vector to the PE, describing its orientation. It is defined as $\mathbf{s} = \mathrm{Im}(\mathbf{B}^* \times \mathbf{B})/ 4 \mu_0 \omega$ \cite{Berry_spin_density}. For a fixed non-zero field amplitude, this vector takes its maximum value when $\varepsilon = 0$.}

\begin{figure}
\centering
\includegraphics[width=7.5cm]{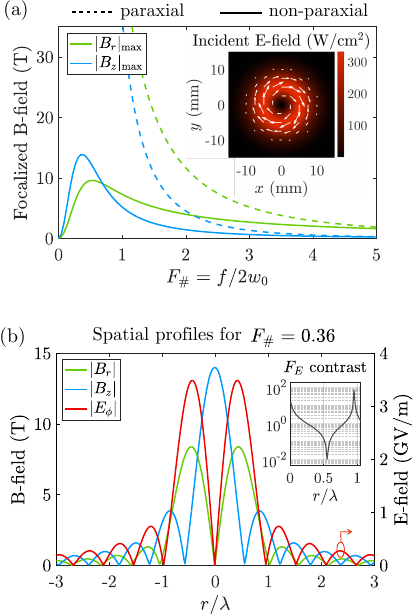}
\caption{Longitudinal B-field enhancement for a single azimuthally polarized laser beam. (a) Maximum $B_r$ (green) and $B_z$ (blue) achieved at focus as a function of the focal ratio, $F_{\#}$, in both paraxial (dashed lines) and non-paraxial (solid lines) formalisms. Inset: transverse intensity (color background) and polarization (arrows) profiles of the driving CVB. (b) Radial profiles of $B_r$ (green), $B_z$ (blue) and $E_{\phi}$ (red) at the focal plane for the optimal focusing conditions, $F_{\#} = 0.36$, in the non-paraxial regime. Inset: contrast $F_E$ between the longitudinal B-field and the E-field.}
\label{single_CVB}
\end{figure}

First, we study the B-field enhancement when focusing an azimuthally polarized beam. Although the results presented in our work could be extended to any visible/infrared laser source, we have considered mid-infrared laser sources ($\lambda = 9 \, \mu \mathrm{m}$), which can reach high intensities in the femtosecond regime \cite{Shumakova:2016vj, Gollner2021}. We have considered $E_0 = 50 \, \mathrm{kV}/\mathrm{m}$, corresponding to a relatively low-intense $3.3 \times 10^2 \, \mathrm{W}/\mathrm{cm}^2$ driving laser beam and $w_0 = 11.25 \, \mathrm{mm}$. The B-field associated to this azimuthal paraxial beam reaches 17~mT for the radial component, while the longitudinal one limits to a very low value of $40$ nT. Fig.~\ref{single_CVB}(a) shows the maximum amplitude of the radial (green) and longitudinal (blue) B-field components obtained for different focal lengths (NA from 1 to 0.1) in the paraxial (dashed lines) and non-paraxial (solid lines) regimes. The transverse intensity and polarization profiles of the driving azimuthally polarized laser beam are presented in the inset of Fig.~\ref{single_CVB}(a). The paraxial solution, given by Eq. (\ref{parax_B}), is expected to provide a good approximation for a focal ratio of $F_{\#} \gg 1$. Indeed, we observe that it fails for $F_{\#}<5$ when describing the radial and longitudinal B-field amplitudes of a tightly focused azimuthally polarized beam as given by Eqs. (\ref{non-parax_Br}) and (\ref{non-parax_Bz}). 
Noticeably, whereas $B_z$ is smaller than $B_r$ in the paraxial regime, this behavior reverses under tight-focusing conditions for $F_{\#} < 0.7$. $B_z$ reaches a maximum value of $14 \, \mathrm{T}$ for $F_{\#} = 0.36$ ($\mathrm{NA} = 0.81$ and $\theta_{\mathrm{max}} = 54^{\circ}$). We shall refer to these values as the optimal focusing conditions.

This situation of maximum $B_z$ enhancement is further analyzed in Fig.~\ref{single_CVB}(b), where we show the radial distribution of $E_{\phi}$ (red), $B_r$ (green) and $B_z$ (blue) when focused with $F_{\#} = 0.36$. As expected, $B_z$ reaches its local maximum at the E-field singularity. We emphasize that a high intensity B-field with an amplitude of 14 Tesla is obtained within a moderately intense azimuthal E-field of $\sim$ 3.5 GV/m, corresponding to an intensity of $1.64 \times 10^{12} \, \mathrm{W}/\mathrm{cm}^2$.
The spatial extent of the B-field isolation from the E-field can be characterized through the contrast, defined as $F_E = c |B_z| / |E_{\phi}|$, and shown in the inset of Fig.~\ref{single_CVB}(b). 

Once we have shown how a tightly focused azimuthally polarized beam provides an enhanced linearly polarized B-field, we explore the control over its polarization state. In particular, we focus our attention in the obtention of circularly polarized B-fields, motivated by their ability to drive a nonlinear response of the magnetization dynamics in ferromagnets \cite{Luis_Sanchez-Tejerina2023}, and by their potentiality to explore chiral mangnetic systems. 
\textcolor{marked_text}{As an initial step towards achieving spatially isolated circularly polarized B-fields, we solve the Maxwell equations to determine the spatial intensity, phase and polarization distributions of the associated E-field. If the spatial B-field distribution is arranged with cylindrical symmetry along the propagation axis, the resulting E-field consists on a vortex beam polarized along the longitudinal direction (see Section 2 of the Supplementary Material). Such fields are exceedingly challenging  to generate and lie beyond our scope.}

\textcolor{marked_text}{Instead, }we propose to combine two azimuthally polarized beams propagating in the $xz$-plane with a crossing angle of $90^{\circ}$ (see beams \#1 and \#2 in Fig.~\ref{focusing_schemes}). The  relative phase delay between the beams is set to $\varphi_{12} = \pi / 2$. \textcolor{marked_text}{Given the configuration considered, the maximum angular aperture of the optic systems compatible with this setup is $\theta_{\mathrm{max}} = 45^{\circ}$ ($F_{\#} = 0.5$ and $\mathrm{NA} = 0.71$). From now on, we will refer to these parameters as the standard focusing conditions. For this situation, $B_z$ reaches $\sim 10$~T,  slightly smaller compared with the optimal focusing conditions, with an azimuthal E-field of 3.2~GV/m ($1.36 \times 10^{12}~\mathrm{W}/\mathrm{cm}^2$).}

\begin{figure}
\centering
\includegraphics[width=8cm]{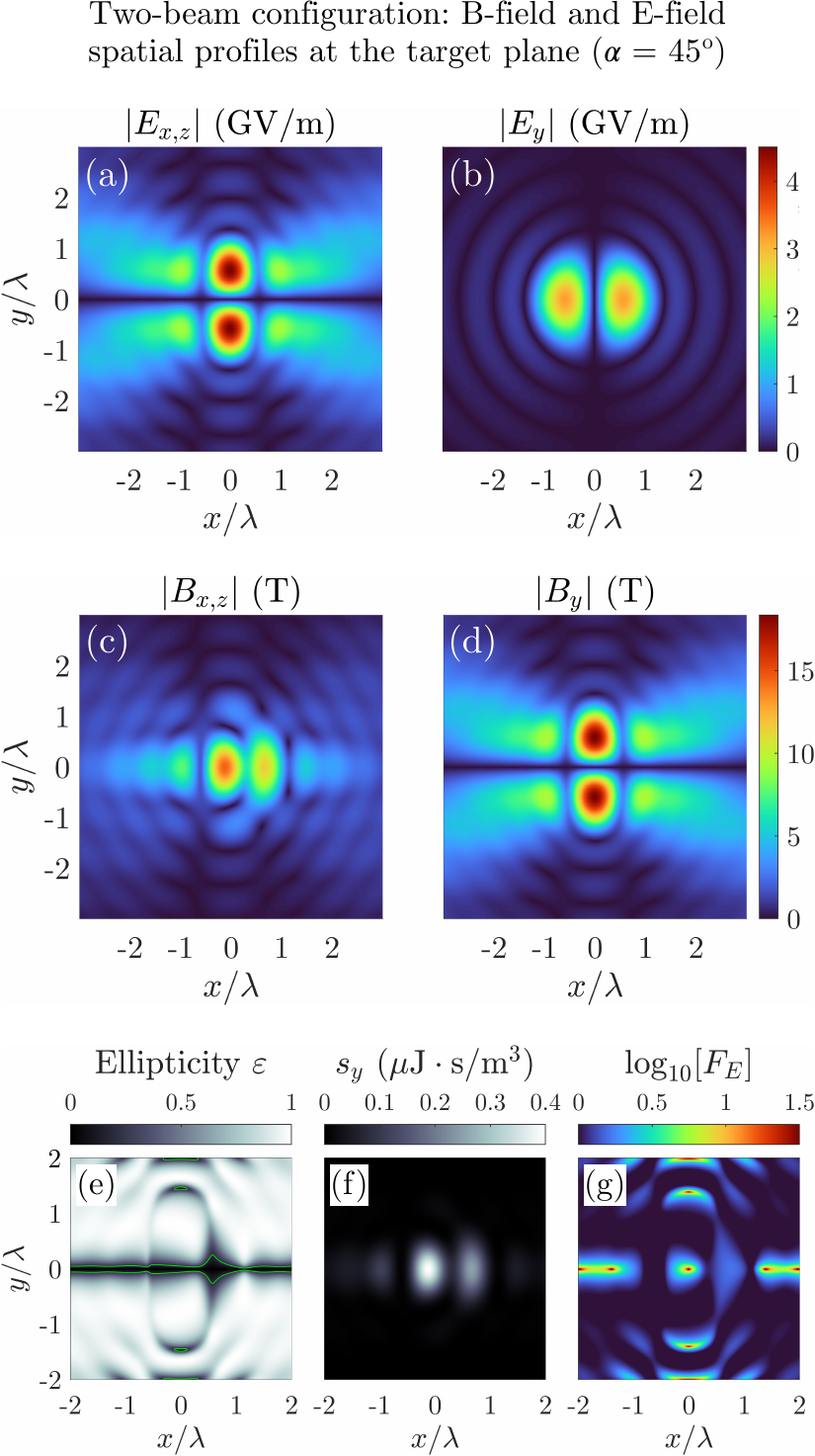}
\caption{Characterization of the field at the target plane for the two-CVB configuration with a relative phase delay of $\pi / 2$. (a)--(d) Spatial distribution of the $E_{x,z}$, $E_y$, $B_{x,z}$ and $B_y$ components respectively. (e) PE ellipticity of the B-field, where the green contour indicate an ellipticity value of $\varepsilon = 0.2$. \textcolor{marked_text}{(f) $y$-component of the spin density $\mathbf{s}$ associated to the B-field. Note that $s_x$ and $s_z$ are negligible in the vicinity of the focal region (not shown).} (g) Spatial distribution of the contrast between the B-field and the E-field, in logarithmic scale.} 
\label{2-beams_plots}
\end{figure}

Figs.~\ref{2-beams_plots}(a)--\ref{2-beams_plots}(d) represent the spatial distribution in the target plane ($xy$-plane, as depicted in Fig. \ref{focusing_schemes}) for an observation angle of $\alpha = 45^{\circ}$ from the $z$-axis of the $E_{x,z}$, $E_{y}$, $B_{x,z}$ and $B_{y}$ components, respectively. 
At the E-field singularity $(x,y)=(0,0)$, two orthogonal magnetic componets with the same spatial distribution are present, $B_x$ and $B_z$, associated to the longitudinal B-field of each beam. The spatial dependent polarization state of the resulting B-field is characterized through the ellipticity, $\varepsilon$, in Fig.~\ref{2-beams_plots}(e), and the $y$-component of the spin density, $s_y$, in Fig.~\ref{2-beams_plots}(f). We identify a spatial needle around $y=0$ where the B-field is circularly polarized ($\varepsilon=0$), extending in the $y$-direction over a width of $\sim 0.1 \lambda = 900 \, \mathrm{nm}$. In such region the PE is mainly oriented along the $y$-direction, $\mathbf{s} \approx (0,s_y,0)$. The spatial distribution of the contrast between the B-field and the E-field, $F_E$, is shown in Fig.~\ref{2-beams_plots}(g). We observe that $\log_{10}\left[F_E\right]>1$ (i.e. the B-field dominates by more than an order of magnitude over the E-field), in a spatial region of radius $\sim 0.14 \lambda = 1.26 \, \mu \mathrm{m}$, considerably larger than that obtained in a single azimuthally polarized beam---see inset of Fig. ~\ref{single_CVB}(b). 


\begin{figure*}[t]
\centering
\includegraphics[width=17.5cm]{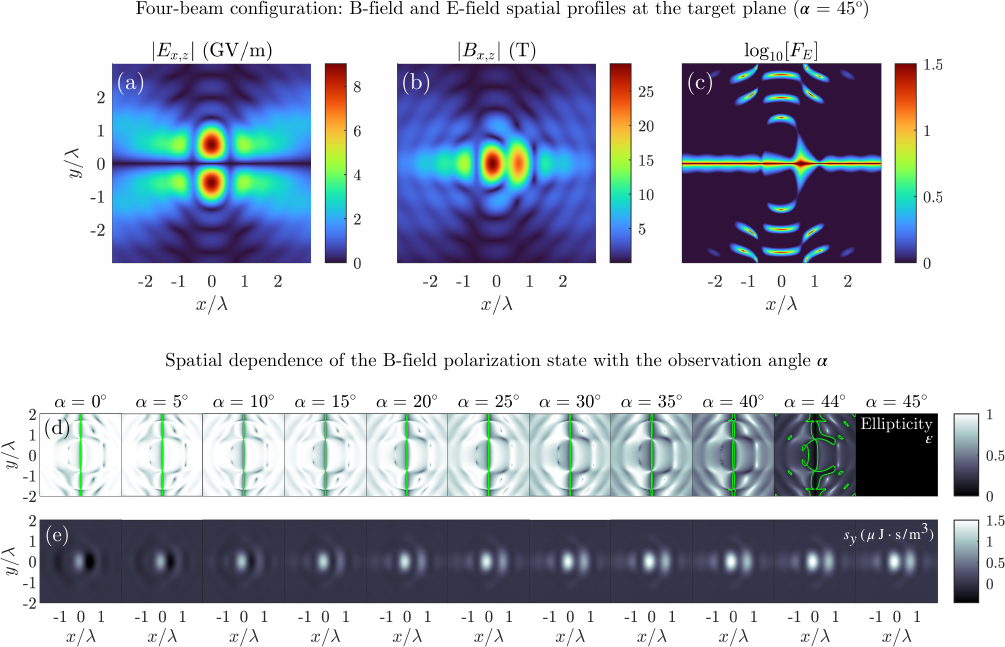}
\caption{Characterization of the field at the target plane for the four-CVB configuration with a relative phase delay of $\pi / 2$ between each consecutive beam. (a)--(c) Spatial distribution of the $E_{x,z}$, $B_{x,z}$ components and the contrast $F_E$, respectively, for the optimum configuration of $\alpha = 45^{\circ}$. \textcolor{marked_text}{(d)--(e) Polarization state of the B-field for different observation angles $\alpha$, through the ellipticity $\varepsilon$ and the $y$-component of the spin density $\mathbf{s}$, respectively ($s_x = s_z = 0$, not shown).} The green contour in (d) indicates the ellipticity value $\varepsilon = 0.2$. Note that the last column panels in (d) and (e) correspond to the case presented in panels (a)--(c), $\alpha = 45^{\circ}$, where an homogeneous circularly polarized B-field polarization state is found at the target plane.}
\label{4-beams_plots}
\end{figure*}

In order to enlarge the spatial region in which the B-field presents an homogeneous polarization state, we propose the coherent addition of four crossing azimuthally polarized beams, as depicted in Fig.~\ref{focusing_schemes}. \textcolor{marked_text}{We consider each beam under the standard tight-focusing condition ($F_{\#} = 0.5$), propagating in the $xz$-plane with a relative crossing angle of $90^{\circ}$.}
The phase delay between each consecutive beam is set to $\pi / 2$, so the relative phase shifts are $\varphi_{1j} = (j-1)\pi/2$ for $j=2,3,4$. Figs.~\ref{4-beams_plots}(a), ~\ref{4-beams_plots}(b) and ~\ref{4-beams_plots}(c) show the spatial distribution of the $E_{x,z}$, $B_{x,z}$ and $F_E$ respectively, for $\alpha = 45^{\circ}$.
Whereas $E_{x,z}$ and $B_{x,z}$ are similar to those of the two-beam configuration, the $y$-components of the fields vanish (not shown) due to the symmetry between the target plane and the four-beam configuration. The first consequence is that the spatial region in which the B-field can be considered to be isolated ($\log_{10}\left[F_E\right]>1$) extends over the $x$-direction.
Second, the absence of $B_y$ results in a homogeneous polarization state in the target plane. This is shown in Figs.~\ref{4-beams_plots}(d) and Figs.~\ref{4-beams_plots}(e), where we plot the spatial distribution of the ellipticity, $\varepsilon$, and the PE direction, $s_y$, respectively, for different \textcolor{marked_text}{observation angles, $\alpha$.} \textcolor{marked_text}{For this setup, $s_x = s_z = 0$ at any observation angle as $B_y=0$.} Thus, in the case of $\alpha= 45^{\circ}$---discussed in Figs. ~\ref{4-beams_plots}(a), ~\ref{4-beams_plots}(b) and ~\ref{4-beams_plots}(c)---the B-field exhibits homogeneous circular polarization state. This characterization at other target plane orientations demonstrate the control that can be achieved over the polarization state of the B-field within that plane.



\begin{figure}[h!]
\centering
\includegraphics[width=8.2cm]{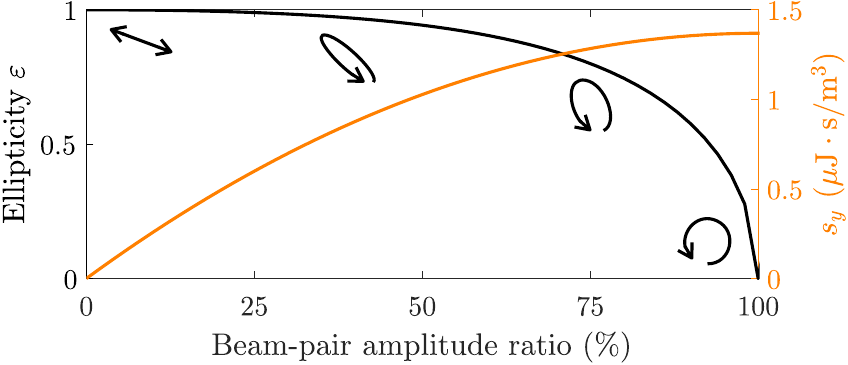}
\caption{B-field polarization state at the focus of the four-CVB configuration in terms of the opposite-pair beams amplitude ratio (CVBs \#1 and \#3 with respect to \#2 and \#4). \textcolor{marked_text}{The PE ellipticity (black) and the $y$-component of the spin density, $s_y$, (orange) show the degree of control achieved over the B-field polarization state at the focus.} Note that $s_x = 0$ and $s_z = 0$ for all amplitude ratios (not shown).}
\label{tunable_exc}
\end{figure}

Further control over the spatial distribution of the polarization state of the isolated B-field can be achieved by tuning the intensity ratio between the driving azimuthally polarized beams, both in the two-beam and four-beam configurations. To illustrate it, we show in Fig.~\ref{tunable_exc} the B-field ellipticity ($\varepsilon$, black) and PE direction ($s_y$, orange) at the focus position $(x,y,z)=(0,0,0)$, when varying the amplitude ratio between opposite beam pairs in the four-beam configuration. When all of the beams exhibit the same amplitude (beam-pair ratio of 100\%), the B-field is circularly polarized, as shown in Fig. \ref{4-beams_plots}. As the beam-pair ratio decreases, the B-field polarization state become elliptical, turning into linear when just two-opposite beams dephased by $\pi$ are present. In addition, the spin density at the focus position always lies in the $y$-direction. It is worth mentioning that for all beam-pair ratios the contrast $F_E$ remains the same as in Fig.~\ref{4-beams_plots}(c), thus the isolation of the B-field is not affected by the tunability of its polarization state.



In conclusion, our study demonstrates the generation of intense and isolated polarization-controlled magnetic fields using tightly focused structured laser beams with azimuthal polarization. We prove that the non-paraxial regime allows for a substantial enhancement of the magnetic longitudinal component carried by an azimuthally polarized laser beam. Such B-field, being linearly polarized and placed at the E-field singularity, can reach tens of Tesla, starting from a moderately intense driving beam (focused intensity of $\sim 10^{12} \, \mathrm{W}/\mathrm{cm}^2$). The implementation of a multiple-beam configuration allows for a precise control of the B-field polarization state, allowing to reach isolated circularly polarized B-fields at the target plane. In addition, this approach also provides further spatial isolation of the B-field. Moreover, we demonstrate that the polarization state of the B-field can be tuned by modifying the amplitude ratio between opposite-pair beams, keeping the isolation with respect to the associated E-field.

The range of potential applications of the presented work is wide within various scientific and technological scenarios, given the importance of achieving full control over selective E-field and B-field excitations \cite{mag_dipole2017,Reynier23}. Applications in probing the ultrafast dynamics of ferromagnetic materials or chiral media have been already mentioned. In addition, the ability of structuring the spatial distribution of the electromagnetic field polarization ushers the possibility to control the magnetization response in a sub-wavelength scale, increasing the potential spatial resolution of optically control magnetization applications. Moreover, if translated into the ultrafast regime, the generation of intense circularly polarized B-field at the femtosecond regime may boost applications in attosecond science \cite{2023_Martin-Hernandez}.

\begin{acknowledgments}
We acknowledge funding from the European Research Council (ERC) under the European Union’s Horizon 2020 research and innovation program (Grant Agreement No.~851201), and from Ministerio de Ciencia e Innovación (PID2022-142340NB-I00).
\end{acknowledgments}

\begin{widetext}

\setcounter{equation}{0}
\setcounter{figure}{0}

\newcommand*\mycommand[1]{\texttt{\emph{#1}}}
\renewcommand{\thefigure}{S\arabic{figure}}
\renewcommand{\theequation}{S\arabic{equation}}

\textcolor{marked_text}{
\section*{Supplementary Material}
The Supplementary Material gives insights into using azimuthally polarized beams to obtain isolated circularly polarized B-fields. Section 1 describes the analytical equations governing the propagation of azimuthally polarized beams in the paraxial approximation. In addition, higher-order vector beams are described and analyzed, demonstrating the suitability of first-order azimuthally polarized beams for generating isolated B-fields. Section 2 solves the Maxwell equations to retrieve the E-field required for obtaining isolated circularly polarized B-fields. The exceedingly complex structure of the retrieved E-fields---a vortex beam polarized along the longitudinal direction---justifies the use of multiple azimuthally polarized beams as a realistic approach to generating isolated circularly polarized B-fields.
}

\subsection*{1. Study of higher-order cylindrical vector beams}

Azimuthally polarized vector beams (AVB) can be decomposed in a superposition of two single-charged vortex beams, with conjugated orbital angular momenta, and opposite circular polarization handedness. For a monochromatic field of frequency $\omega~=~2 \pi c / \lambda$ and wavelength $\lambda$ in cylindrical coordinates $(r,\phi,z)$ propagating along the $z$-axis:
\begin{gather}
    \mathbf{E}_{\text{AVB}}(\mathbf{r},t) = E_0\frac{r}{w_0}e^{-\frac{r^2}{w_0^2}}e^{i(\frac{2\pi}{\lambda}z-\omega t)}\frac{1}{\sqrt{2}}\left[e^{-i\phi}\hat{\mathbf{u}}_{\text{R}} - e^{i\phi}\hat{\mathbf{u}}_{\text{L}} \right],
\end{gather}
where $\hat{\mathbf{u}}_{\text{R}} = (\hat{\mathbf{u}}_{x} + i \hat{\mathbf{u}}_{y})/\sqrt{2}$ and $\hat{\mathbf{u}}_{\text{L}} = (\hat{\mathbf{u}}_{x} - i \hat{\mathbf{u}}_{y})/\sqrt{2}$ are the right and left circular polarization unit vectors, $E_0$ is the amplitude and $w_0$ is the beam waist. One could think about a more general, high-order vector beam (VB) wihtout restricting the orbital angular momentum (OAM) value of the conjugated  vortices to $\ell=\pm 1$. For an arbitrary $| \ell | \geq 1$ value, the VB is given as
\begin{gather}
    \mathbf{E}_{\mathrm{VB}}(\mathbf{r},t) = E_0\left(\frac{r}{w_0}\right)^{|\ell|}e^{-\frac{r^2}{w_0^2}}e^{i(\frac{2\pi}{\lambda}z-\omega t)}\frac{1}{\sqrt{2}}\left[e^{-i\ell\phi}\hat{\mathbf{u}}_{\text{R}} - e^{i\ell\phi}\hat{\mathbf{u}}_{\text{L}} \right].
\end{gather}
Expanding and operating within the last bracket, the VB can be expressed in terms of the radial and azimuthal components perpendicular to the propagation direction,
\begin{gather}
    \mathbf{E}_\perp(\mathbf{r},t) = E_r(\mathbf{r},t)\hat{\mathbf{u}}_r + E_\phi(\mathbf{r},t)\hat{\mathbf{u}}_\phi\\
    E_r(\mathbf{r},t) = E_0\left(\frac{r}{w_0}\right)^{|\ell|}e^{-\frac{r^2}{w_0^2}}e^{i(\frac{2\pi}{\lambda}z-\omega t + \frac{\pi}{2})}\sin\left[(1-\ell)\phi\right]\\
    E_\phi(\mathbf{r},t) = E_0\left(\frac{r}{w_0}\right)^{|\ell|}e^{-\frac{r^2}{w_0^2}}e^{i(\frac{2\pi}{\lambda}z-\omega t + \frac{\pi}{2})}\cos\left[(1-\ell)\phi\right],
\end{gather}
where $\hat{\mathbf{u}}_r = \cos\phi \, \hat{\mathbf{u}}_x + \sin\phi \, \hat{\mathbf{u}}_y$ and $\hat{\mathbf{u}}_\phi = -\sin\phi \, \hat{\mathbf{u}}_x + \cos\phi \, \hat{\mathbf{u}}_y$ are the radial and azimuthal polarization unit vectors respectively. %
Calculating the divergence of the electric field (E-field) $\nabla \cdot \mathbf{E} = \frac{1}{r}\partial_r(rE_r) + \frac{1}{r}\partial_\phi E_\phi + \partial_z E_z$ and being
\begin{align}
    \frac{1}{r}\partial_r(rE_r) &= i\frac{E_0}{r}\left(\frac{r}{w_0}\right)^{|\ell|}e^{-\frac{r^2}{w_0^2}}\sin\left[(1-\ell)\phi\right]\left[\ell+1-\frac{2r^2}{w_0^2}\right]e^{i(\frac{2\pi}{\lambda}z-\omega t)},\\
    \frac{1}{r}\partial_\phi E_\phi &= -i\frac{E_0}{r}\left(\frac{r}{w_0}\right)^{|\ell|}e^{-\frac{r^2 }{w_0^2}}\left(1-\ell\right)\sin\left[(1-\ell)\phi\right]e^{i(\frac{2\pi}{\lambda}z-\omega t)};
\end{align}
a longitudinal E-field component $E_z$ is needed to satisfy Maxwell equations. Assuming $\partial_z E_z = i\frac{2\pi}{\lambda} E_z(r,\phi)$, the required $E_z$ component is
\begin{align}
    E_z(\mathbf{r},t) &= \frac{i}{2\pi r/\lambda}\left[\partial_r(rE_r) + \partial_\phi E_\phi \right] \nonumber \\ 
    &= -\frac{E_0}{\pi w_0/\lambda}\left(\frac{r}{w_0}\right)^{|\ell|-1}e^{-\frac{r^2}{w_0^2}}\sin\left[(1-\ell)\phi\right]\left[\ell + \frac{r^2}{w_0^2}\right]e^{i(\frac{2\pi}{\lambda}z-\omega t)}.
\end{align}
The longitudinal E-field $E_z$ amplitude is $\pi w_0/\lambda$ times lower than the radial and azimuthal components. For example, for a wavelength $\lambda = 0.8\:\mathrm{\mu m}$ and a $w_0 = 3\:\mathrm{\mu m}$ waist, the longitudinal E-field peak amplitude is $\sim 0.08$ times weaker than the other two components. Thus, the longitudinal E-field can be neglected in a first approximation. In the particular case of $|\ell|=1$, i.e. an AVB, the radial and azimuthal derivatives cancel out as both are proportional to $\sin\left[(1-\ell)\phi\right]$, and the longitudinal E-field component is exactly zero.

The magnetic field (B-field) components can be derived from the second Maxwell equation, $\nabla\times\mathbf{E}~=~\partial_t \mathbf{B}$. The right hand side is trivially calculated as $\partial_t \mathbf{B}(\mathbf{r},t) = i\omega \mathbf{B}(\mathbf{r},t)$. In cylindrical coordinates the rotational is
\begin{align}
    \label{eq:Efieldrot}
    \nabla \times \mathbf{E} = \underbrace{\left[\frac{1}{r}\partial_\phi E_z - \partial_z E_\phi\right]}_{i\omega B_r}\hat{\mathbf{u}}_r + \underbrace{\left[\partial_z E_r - \partial_r E_z\right]}_{i\omega B_\phi}\hat{\mathbf{u}}_\phi + \underbrace{\frac{1}{r}\left[\partial_r(rE_\phi) - \partial_\phi E_r\right]}_{i\omega B_z}\hat{\mathbf{u}}_z.
\end{align}
Using the loose focusing approximation $w_0\gg\lambda$, the longitudinal E-field can be neglected. Note that the $z$-derivatives trivially evaluate to $\partial_z E_r= i\frac{2\pi}{\lambda} E_r$ and $\partial_z E_\phi = i\frac{2\pi}{\lambda}E_\phi$. The terms in the last bracket in Eq. (\ref{eq:Efieldrot}) are
\begin{align}
    &\frac{1}{r}\partial_r(rE_\phi) = i\frac{E_0}{w_0}\left(\frac{r}{w_0}\right)^{|\ell|-1}e^{-\frac{r^2}{w_0^2}}\cos\left[(1-\ell)\phi\right]\left[l+1 - \frac{2r^2}{w_0^2}\right]e^{i(\frac{2\pi}{\lambda}z-\omega t)},\\
    &\frac{1}{r}\partial_\phi E_r = i\frac{E}{w_0}\left(\frac{r}{w_0}\right)^{|\ell|-1}e^{-\frac{r^2}{w_0^2}}(1-\ell)\cos\left[(1-\ell)\phi\right]e^{i(\frac{2\pi}{\lambda}z-\omega t)}.
\end{align}

Within the used approximations, the B-field components are given by the following expressions:
\begin{align}
    &B_r(\mathbf{r},t) = -\frac{E_0}{c}\left(\frac{r}{w_0}\right)^{|\ell|}e^{-\frac{r^2}{w_0^2}}\cos\left[(1-\ell)\phi\right]e^{i(\frac{2\pi}{\lambda}z-\omega t - \frac{\pi}{2})},\\
    &B_\phi(\mathbf{r},t) = \frac{E_0}{c}\left(\frac{r}{w_0}\right)^{|\ell|}e^{-\frac{r^2}{w_0^2}}\sin\left[(1-\ell)\phi\right]e^{i(\frac{2\pi}{\lambda}z-\omega t + \frac{\pi}{2})},\\
    &B_z(\mathbf{r},t) = 2 \frac{E_0}{\omega w_0}\left(\frac{r}{w_0}\right)^{|\ell|-1}e^{-\frac{r^2}{w_0^2}}\cos\left[(1-\ell)\phi\right]\left[\ell-\frac{r^2}{w_0^2}\right]e^{i(\frac{2\pi}{\lambda}z - \omega t)},
\end{align}
where it has been used $\lambda \omega = 2\pi c$. Note that for an AVB, $|\ell|=1$, we recover the same expression for the B-field shown in Eq. (2) in  the main text. As the longitudinal B-field component is now proportional to $r^{|\ell|-1}$, the $B_z$ intensiy distribution at $r=0$ has a singularity for all values of $\ell$, except in the case of an AVB. Moreover, the term $\cos\left[(1-\ell)\phi\right]$ structures the intensity distribution into a necklace-like profile along the azimuthal direction with opposite phases between contigous lobes, not being isolated from the associated E-field. In Fig. \ref{fig:HCVB} we present the transverse E-field (top row), transverse B-field (middle row) and longitudinal B-field (bottom row) inentsity and polarization distributions for $\ell=1,3,5$ values.

As it can be observed in Fig. \ref{fig:HCVB}, the use of higher-order modes does not allow for spatial isolation of the B-field from the E-field. Thus, we restrict our study to the use of AVBs.

\begin{figure*}[h!]
    \centering
    \includegraphics[width=0.75\textwidth]{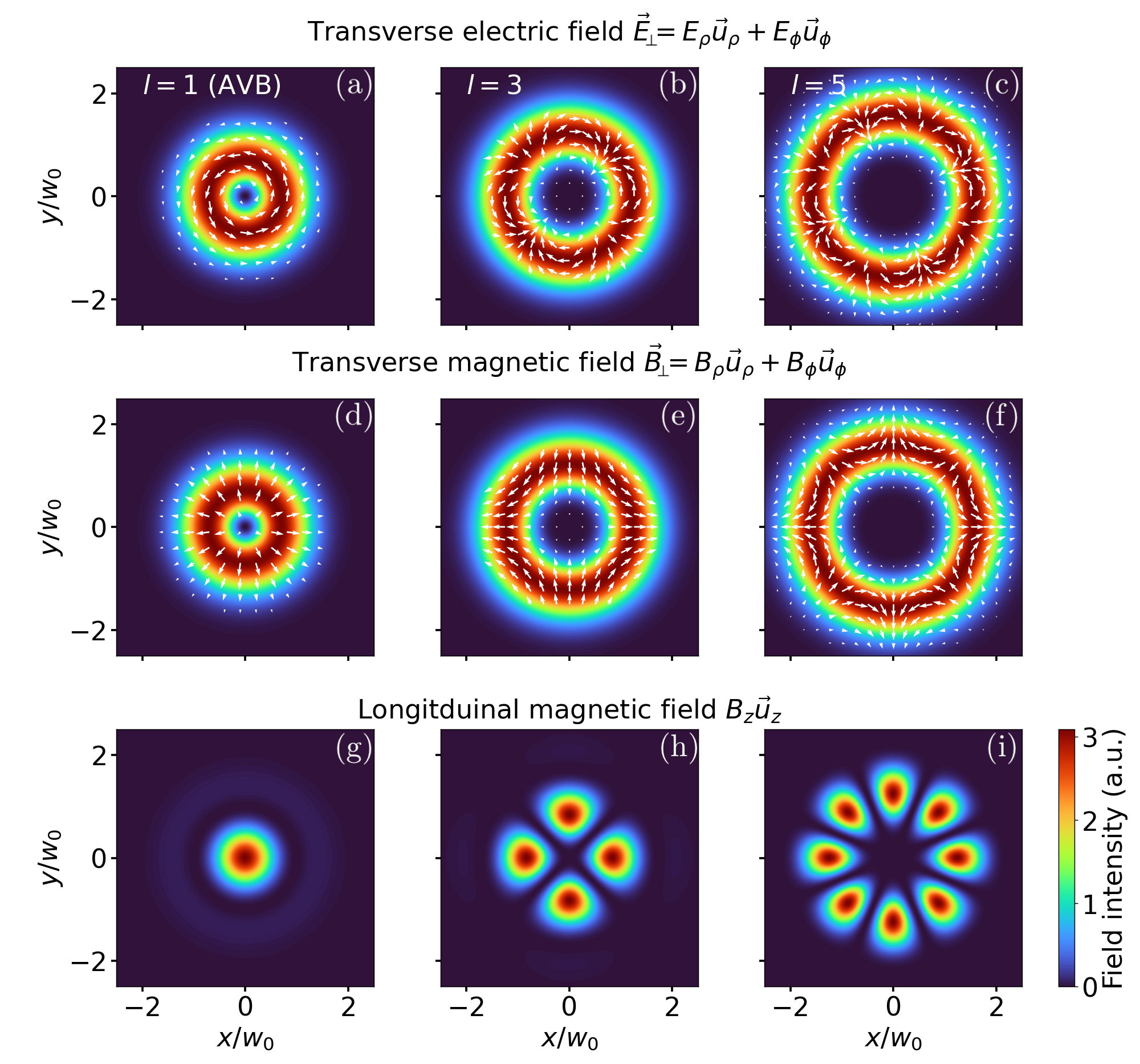}
    \caption{Transverse intensity distributions of the vector beams for OAM values $\ell=1,3,5$ (consecutive columns from left to right) in the loose focusing approximation $w_0 \gg \lambda$. Arrows indicate the local polarization state. (a)--(c) Transverse E-field. (d)--(f) Transverse B-field. (g)--(i) Longitudinal B-field. For $|\ell| > 1$ there is no longitudinal B-field component at the beam axis as in the case of the AVB shown in (g).}
    \label{fig:HCVB}
\end{figure*}


\subsection*{2. Prediction from the numerical solution of the Maxwell equations to obtain isolated B-fields}

In this section we solve numerically the Maxwell equations to give insights into the electromagnetic field chosen to obtain spatially isolated circularly polarized B-fields.

Following the axes criteria of the main text, the desired B-field is composed of two orthogonal components $B_x$ and $B_z$ dephased by $\pi/2$, lying in the $xz$-plane. For simplicity, we model here their spatial distribution as a Gaussian transverse profile with cylindrical symmetry along the propagation $y$-axis, $|B_{x,z}(\mathbf{r})| \propto e^{-(x^2 + z^2) / w_0^2}$. Additionally, $B_z$ is multiplied by a phase factor of $e^{i \pi /2}$. Once this circularly polarized B-field is designed, we solve numerically the Maxwell equations to determine the spatial intensity, phase and polarization distributions of the associated E-field. In particular we use $\nabla \times \mathbf{B} = -i \omega \mathbf{E} /c^2$ (monochromatic field), ensuring that $\nabla \cdot \mathbf{E} = 0$. 

Fig. \ref{fig:ME_circular_Bfield} shows the intensity, polarization and phase spatial profiles of the resulting E-field and B-field at the $xz$-plane. Circles and crosses give the local polarization state, corresponding to circularly and linearly polarized states, respectively. As we can observe, the resulting E-field consists on a vortex beam exhibiting linear polarization along the longitudinal---propagation---direction. 
\begin{figure*}[h!]
    \centering
    \vspace{2ex}
    \includegraphics[width=0.75\textwidth]{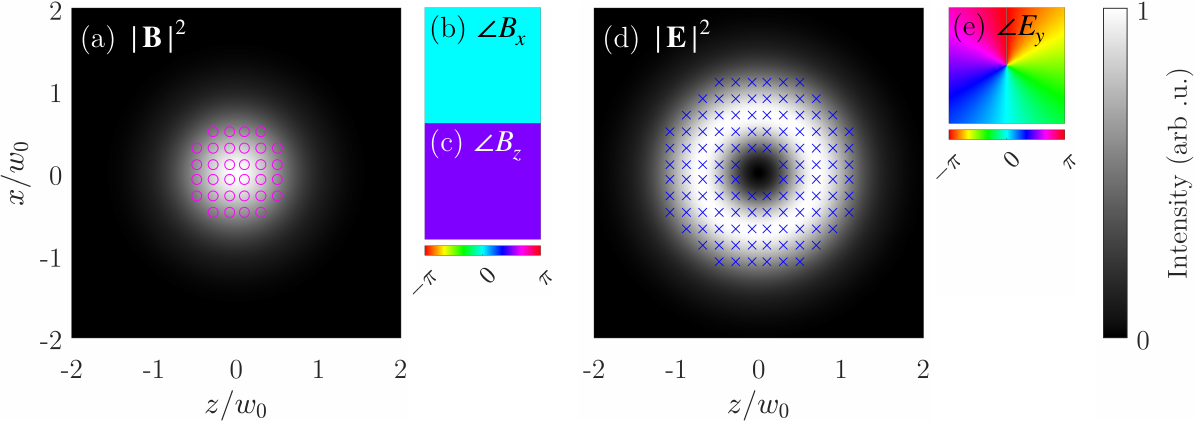}
    \caption{(a)--(c) Transverse intensity and phase distribution of an initial circularly polarized B-field propagating along the $y$-direction. (d) Associated E-field, obtained through numerical solution of the Maxwell equations. The inset (e) shows that the phase profile of $E_y$ corresponds to that of a vortex beam with topological charge of one. The homogeneous polarization states of the fields are represented by circles and crosses (i.e. circular and linear polarization states respectively).}
    \label{fig:ME_circular_Bfield}
\end{figure*}

Such electromagnetic field is exceedingly challenging to generate. To circumvent this drawback, we superimpose several tightly-focused azimuthally polarized vector beams to obtain isolated circularly polarized B-fields, as explained in the main text.
Note that such configurations (either the use of two or four beams), satisfy the solution of the Maxwell equations.  

\end{widetext}

\bibliography{bibliography}

\end{document}